\documentclass[runningheads]{llncs}
\usepackage{algorithm}
\usepackage{algpseudocode}
\usepackage{multirow}

\usepackage{amssymb}
\usepackage{bbding}
\usepackage{xcolor,colortbl}
\definecolor{Gray}{gray}{0.85}
\newcolumntype{a}{>{\columncolor{Gray}}l}
\usepackage[T1]{fontenc}
\usepackage{graphicx}
\def\Fig#1{{Fig.~\ref{fig:#1}}}

\def\Table#1{{Table~\ref{tbl:#1}}}

\mathchardef\mhyphen="2D

\begin{document}
\title{Liver Tumor Screening and Diagnosis in CT with Pixel-Lesion-Patient Network\thanks{Partially supported by the National Natural Science Foundation of China (grant 82071885), Basic Research Projects of Liaoning Provincial Department of Education (LJKMZ20221160), the National Youth Talent Support Program of China, and Science and Technology Innovation Talent Project in Shenyang (RC210265).}}
\titlerunning{Liver Tumor Screening and Diagnosis in CT}
%
\author{Ke Yan\inst{1,2} \Envelope \and 
Xiaoli Yin\inst{3} \and
Yingda Xia\inst{1} \and
Fakai Wang\inst{1} \and
Shu Wang\inst{3} \and \\
Yuan Gao\inst{1,2} \and
Jiawen Yao\inst{1,2} \and
Chunli Li\inst{1,2,3} \and
Xiaoyu Bai\inst{1,2} \and \\
Jingren Zhou\inst{1,2} \and
Ling Zhang\inst{1} \and
Le Lu\inst{1} \and
Yu Shi\inst{3} \Envelope }

\institute{DAMO Academy, Alibaba Group \and
Hupan Lab, 310023, Hangzhou, China \and
Department of Radiology, Shengjing Hospital of China Medical University, Shenyang 110004, China\\
yanke.yan@alibaba-inc.com; 18940259980@163.com
}
\authorrunning{K. Yan, X. Yin et al.}
%

%
\maketitle              
\begin{abstract}
Liver tumor segmentation and classification are important tasks in computer aided diagnosis. We aim to address three problems: liver tumor screening and preliminary diagnosis in non-contrast computed tomography (CT), and differential diagnosis in dynamic contrast-enhanced CT. A novel framework named Pixel-Lesion-pAtient Network (PLAN) is proposed. It uses a mask transformer to jointly segment and classify each lesion with improved anchor queries and a foreground-enhanced sampling loss. 
It also has an image-wise classifier to effectively aggregate global information and predict patient-level diagnosis. A large-scale multi-phase dataset is collected containing 939 tumor patients and 810 normal subjects. 4010 tumor instances of eight types are extensively annotated. On the non-contrast tumor screening task, PLAN achieves 95\% and 96\% in patient-level sensitivity and specificity. On contrast-enhanced CT, our lesion-level detection precision, recall, and classification accuracy are 92\%, 89\%, and 86\%, outperforming widely used CNN and transformers for lesion segmentation. We also conduct a reader study on a holdout set of 250 cases. PLAN is on par with a senior human radiologist, showing the clinical significance of our results.

\end{abstract}

\section{Introduction}

Liver cancer is the third leading cause of cancer death world-wide in 2020~\cite{Sung2021GLOBOCAN}. Early detection and accurate diagnosis of liver tumors may improve overall patient outcomes, in which imaging plays a key role~\cite{Marrero2014ACG}. Computed tomography (CT) is one of the most important imaging modalities for liver tumors. Dynamic contrast-enhanced (DCE) CT is widely used for diagnostics, but it requires iodine contrast injection which can cause reaction and potential risks in patients. Recently, non-contrast (NC) CT scans are gaining attention as they are cheaper and safer to acquire, thus can be potential tools for opportunistic tumor screening~\cite{Xia2021pancreas,Yao2022eso}. Meanwhile, finding and diagnosing tumors in NC CTs is also extremely challenging because of the poor contrast between tumors and normal tissues compared to those in DCE CTs. Prior works on pancreas~\cite{Xia2021pancreas} and esophagus~\cite{Yao2022eso} have shown that latest deep learning techniques can detect subtle texture and shape changes in NC CT that even human eyes may miss. Thus, we aim to investigate the performance of liver tumor segmentation and classification in NC CTs. Such an approach will be helpful to discover asymptomatic incidental tumors~\cite{Semaan2016incident} from routine NC CT scans indicated for general diagnostic purposes at no additional cost and radiation exposure. 
After an incidental tumor is found, the patient may undergo further imaging examination such as a multi-phase DCE CT for differential diagnosis~\cite{Marrero2014ACG}, which can provide useful discriminative information such as the vascularity of lesions and the pattern of contrast agent enhancement~\cite{Xu2023class}. Liver is largest solid organ in body and is the site of many tumor types~\cite{Marrero2014ACG}. Therefore, accurate tumor type classification is important for the decision of treatment plans and prognosis. 

Many researchers have developed algorithms to automatically segment~\cite{Bilic2019LiTS,Li2018HDense,Seo2020mUNet,Tang2020E2Net,Zhang2021modal} or classify~\cite{Zhou2021Frontiers,Yasaka2018differ,Xu2023class} liver tumors in CT to help radiologists improve their accuracy and efficiency. For example, public datasets such as the Liver Tumor Segmentation Benchmark (LiTS)~\cite{Bilic2019LiTS} fostered a series of works aiming to segment liver tumors with improved convolutional neural network (CNN) backbones~\cite{Li2018HDense,Seo2020mUNet} and lesion edge information~\cite{Tang2020E2Net}. LiTS only has single-phase CTs (venous phase). Several studies investigated methods to exploit multi-phase CT by methods such as hetero-phase fusion~\cite{Cheng2022HCC} and modality-aware mutual learning~\cite{Zhang2021modal}. 
There are few work discussing liver tumor analysis in NC CT~\cite{Cheng2022HCC}. Besides lesion segmentation, CNN-based lesion classification algorithms have been studied to distinguish common lesion types~\cite{Xu2023class,Zhou2021Frontiers,Yasaka2018differ}.

In this paper, we build a comprehensive framework to address both tumor screening and diagnosis
. (1) Tumor screening involves finding tumor patients in a large pool of healthy subjects and patients. Most existing works in tumor segmentation and detection did not explicitly consider it since their training and testing images are all tumor patients. Such models may generate false positives in real-world screening scenario when facing diverse tumor-free images. We collect a large-scale dataset with both tumor and non-tumor subjects, where the non-tumor subjects includes not only healthy ones, but also patients with various diffuse liver diseases such as steatosis and hepatitis to improve the robustness of the algorithm. (2) Most works studied liver tumor segmentation alone without differentiating tumor types, while a few works classify liver tumors on cropped tumor patches~\cite{Xu2023class,Zhou2021Frontiers,Yasaka2018differ}. Meanwhile, we learn tumor segmentation and classification with one network using an instance segmentation framework~\cite{Cheng2022Mask2Former}. 
We train two networks for NC and multi-phase DCE CTs, respectively. (3) For evaluation, previous segmentation works typically use \emph{pixel-level} metrics such as Dice coefficient. Such metrics cannot reflect the \emph{lesion-level} accuracy (how many lesion instances are correctly detected and classified) and may bias to large lesions when a patient has multiple tumors. \emph{Patient-level} metrics (e.g. classifying whether a subject has malignant tumors) are also useful for treatment recommendation in clinical practice~\cite{Xia2021pancreas,Yao2022eso}. Therefore, we assess our algorithm thoroughly with pixel, lesion, and patient-level metrics.

Algorithms for liver tumor segmentation have focused on improving the feature extraction backbone of a fully-convolutional CNN~\cite{Li2018HDense,Seo2020mUNet,Tang2020E2Net,Zhang2021modal}. The pixel-wise segmentation architectures may not be optimal for lesion and patient-level evaluation metrics since they cannot consider a lesion or an image holistically. Recently, a series of mask transformer algorithms~\cite{Wang2021MaxDeepLab,Cheng2021MaskFormer,Cheng2022Mask2Former} have emerged in the computer vision community and achieved the state-of-the-art performance in instance segmentation tasks. In brief, they use object queries to interact with image feature maps and with each other to produce mask and class predictions for each instance. Inspired by them, we propose a novel end-to-end framework named Pixel-Lesion-pAtient Network (PLAN) for lesion segmentation and classification, as well as patient classification. It contains three branches with bottom-up cooperation: The segmentation map from the \emph{pixel branch} helps to initialize the \emph{lesion branch}, which is an improved mask transformer aiming to segment and classify each lesion; The \emph{patient branch} aggregates information from the whole image and predicts image-level labels of each lesion type, with regularization terms to encourage consistency with the lesion branch.

We collected a large-scale multi-phase dataset containing 810 non-tumor subjects and 939 tumor patients. 4010 tumor instances of eight types are extensively annotated based on pathological reports. On the non-contrast tumor screening and diagnosis task, PLAN achieves 95.0\%, 96.4\%, and 0.965 in patient-level sensitivity, specificity, and average AUC for malignant and benign patients, in contrast to 94.4\%, 93.7\%, and 0.889 for the widely-used nnU-Net~\cite{Isensee2021nnUNet}. On multi-phase DCE CT, our lesion-level detection precision, recall, and classification accuracy are 92.2\%, 89.0\%, 85.9\%, outperforming nnU-Net~\cite{Isensee2021nnUNet} and Mask2Former~\cite{Cheng2022Mask2Former}. We further conduct a reader study on a holdout set of 250 cases. Our algorithm is on par with a senior radiologist (16 yrs experience), showing the clinical significance of our results. Our codes will be made public upon institutional approval.

\section{Method}

\subsection{Preliminary on Mask Transformer}

\begin{figure}[t]
\centering
\includegraphics[width=1\textwidth,trim=0 120 250 0, clip]{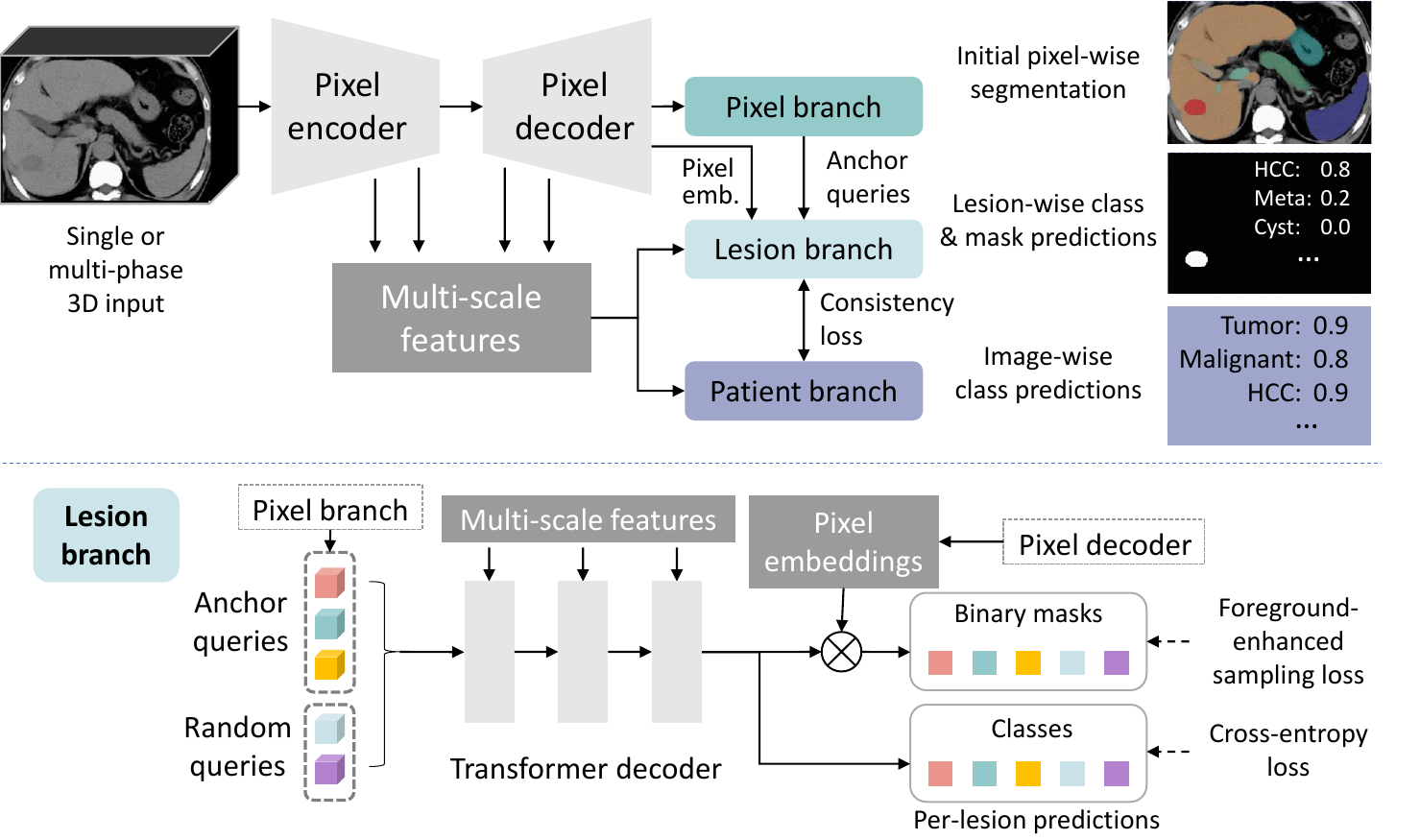}
\caption{Framework of the Pixel-Lesion-pAtient Network (PLAN).} \label{fig:framework} 
\end{figure}

Mask transformers are a series of latest works achieving superior accuracy on various segmentation tasks~\cite{Wang2021MaxDeepLab,Cheng2021MaskFormer,Cheng2022Mask2Former,Yu2022KMax}. Different from traditional fully-convolutional segmentators~\cite{Isensee2021nnUNet} that predict a class label for each pixel, mask transformers predict a class label and a binary mask for each object
. Take Mask2Former~\cite{Cheng2022Mask2Former} as an example. It includes a pixel encoder and a pixel decoder that extract a high-resolution pixel embedding tensor $\mathbf{P}\in \mathbb{R}^{M\times D\times H\times W}$ from the image, where $M$ is the embedding dimension, 
 $D\times H\times W$ is the shape of the 3D image. 
 A group of $Q$ learnable feature vectors $\{\mathbf{q}_i\in \mathbb{R}^{M}\}_{i=1}^{Q}$ are randomly initialized as object queries. They are processed by a transformer decoder to interact with multi-scale image features and each other using cross and self-attention operations. After processing, each query is supposed to contain information of one object, which can be used to predict the class probability $\mathbf{c}\in \mathbb{R}^{C+1}$ of the object. Here $C$ is the number of object classes, and we add 1 to indicate an additional ``no-object'' class if the query does not match with any object. In training, Mask2Former uses bipartite matching~\cite{Carion2020DETR} to assign each query to a ground-truth object (or ``no-object''). Multiplying $\mathbf{q}_i$ with $\mathbf{P}$ gives the binary mask $\mathbf{m}_i\in \mathbb{R}^{D\times H\times W}$ of object $i$. During inference, the class and mask predictions of all queries can be merged by matrix multiplication to obtain the final semantic segmentation result $\hat{\mathbf{Y}}\in\mathbb{R}^{C\times D\times H\times W}$. 
 We refer readers to~\cite{Cheng2022Mask2Former} for more details.

Mask transformers have various advantages when applied to our task. They can classify a lesion as a whole instead of classifying each pixel, thus can view each lesion holistically. Cross-attention is used to aggregate global features for each lesion. Inter-lesion relation can also be exploited by self-attention operations. In liver CT, inter-lesion relation is diagnostically useful, e.g., metastases and cysts are often multiple. Therefore, We pioneer mask transformers' adaptation for lesion segmentation and classification in 3D medical images. Given a ground-truth or a predicted lesion mask image, we perform connected component (CC) analysis and treat each CC as a lesion instance for training and evaluation.

\subsection{Pixel-Lesion-Patient Network (PLAN)}

Our goal is to segment the mask and classify the type of each tumor in a liver CT. We also hope to make patient-level diagnoses for each CT scan. PLAN is inspired by Mask2Former~\cite{Cheng2022Mask2Former} with three key improvements: (1) A pixel branch is added to provide anchor queries to the lesion branch. (2) The lesion branch is composed of the transformer decoder in Mask2Former, and we improve its segmentation loss to enhance recall of small lesions. (3) A patient branch is attached to make dedicated image-level predictions with a proposed lesion-patient consistency loss. 
Our framework is shown in \Fig{framework}.

{\bf Pixel branch and anchor queries.} The pixel branch is a convolutional layer after the pixel decoder and 
learns to predict pixel-wise segmentation maps similar to traditional segmentators. 
We do CC analysis to the predicted mask to extract lesion instances, and then average the pixel embeddings inside each predicted lesion to obtain a feature vector. The feature vectors are regarded as anchor queries and work the same way as the randomly initialized queries in the lesion branch. Compared to the random queries in the original Mask2Former, the anchor queries contain prior information of the lesions to be segmented, helping the lesion branch to match with the lesion targets more easily~\cite{Liu2022DAB}. 

{\bf Lesion branch and foreground-enhanced sampling loss.} Similar to Mask2Former, the lesion branch predicts a binary mask and a class label for each query, see \Fig{framework}. 
Mask2Former calculates its segmentation loss on $K$ sampled pixels instead of on the whole image, which is shown to both improve accuracy and reduce GPU memory usage~\cite{Cheng2022Mask2Former}. However, in lesion segmentation, some tumors are very small compared to the whole 3D image. The importance sampling strategy~\cite{Cheng2022Mask2Former} can hardly select any foreground pixels in such cases, so the loss only contains background pixels, degrading the segmentation recall of small lesions. We propose a simple approach to remedy this issue by sampling an extra $n$ foreground pixels for each lesion.  

{\bf Patient branch.} A patient-level diagnosis is useful for triage. For example, diagnosing the subject as normal, benign, or malignant will result in completely different treatments~\cite{Zhao2021pancreas}. Intuitively, we can also infer patient-level labels from segmentation results by checking if there is any lesion in the predicted mask. However, certain tumors are often related to signs outside the tumor, e.g. hepatocellular carcinoma and cirrhosis, cholangiocarcinoma and bile duct dilatation, etc. We equip PLAN with a dedicated patient branch to aggregate such global information to make better patient-level prediction
. Since one patient can have multiple liver tumors of different types, in our problem, we give each image several hierarchical binary labels. The first label classifies normal and tumor subjects (whether the image contains any tumor); The second and third labels indicate the existence of respectively benign and malignant tumors; The rest $C$ labels suggest the existence of $C$ fine-grained types of tumors. 
We employ the dual-path transformer block~\cite{Wang2021MaxDeepLab} to fuse multi-scale features from the pixel encoder and decoder to generate a feature map, followed by global average pooling and a linear classification layer to predict the $C+3$ labels. 

A {\bf lesion-patient consistency loss} is further proposed to encourage coherence of the lesion and patient-level predictions. Inspired by multi-instance learning~\cite{Cheplygina2019survey}, we compute a pseudo patient-level prediction $\tilde{\mathbf{c}}\in \mathbb{R}^{C}$ from the lesion-level predictions by max-pooling the class probability of each class across all lesion queries (discarding the no-object class). We also have the probability vector from the patient branch $\tilde{\mathbf{p}}\in \mathbb{R}^{C}$ corresponding to the $C$ fine-grained classes. Then, we compute the L2 loss between them: $\mathcal{L}_{\mathrm{consist}}=\|\tilde{\mathbf{p}} - \tilde{\mathbf{c}}\|^2$. 

The overall loss of PLAN is listed in Eq.~\ref{eq:loss}, where $\mathcal{L}_\mathrm{pixel}$ is the combined cross-entropy (CE) and Dice loss for the pixel branch as in nnU-Net~\cite{Isensee2021nnUNet}; $\mathcal{L}_{\mathrm{lesion\mhyphen class}}$ is the CE loss~\cite{Cheng2022Mask2Former} for lesion classification in the lesion branch; $\mathcal{L}_\mathrm{lesion\mhyphen mask}$ is the combined CE and Dice loss~\cite{Cheng2022Mask2Former} for binary lesion segmentation in the lesion branch with the foreground-enhanced sampling strategy; $\mathcal{L}_\mathrm{patient}$ is the binary CE loss for the multi-label classification task in the patient branch.
\begin{equation}
\label{eq:loss}
    \mathcal{L}=\lambda_1\mathcal{L}_\mathrm{pixel} + \lambda_{2\mathrm{c}}\mathcal{L}_{\mathrm{lesion\mhyphen class}} + \lambda_{2\mathrm{m}}\mathcal{L}_\mathrm{lesion\mhyphen mask} + \lambda_3\mathcal{L}_\mathrm{patient} + \lambda_4\mathcal{L}_\mathrm{consist}.
\end{equation}

\section{Experiments}

{\bf Data.} Our dataset contains 810 normal subjects and 939 patients with liver tumors
. Each normal subject has a non-contrast (NC) CT, while each patient has a dynamic contrast-enhanced (DCE) CT scan with NC, arterial, and venous phases. We use DEEDS~\cite{Heinrich2013DEEDS} to register NC and arterial phases to the venous phase, and then invite a senior radiologist with 10 years of experience to annotate on the multi-phase CTs using CT Labeler~\cite{Wang2023CTLabeler}. The 3D mask and the type of all liver tumors are annotated based on pathological reports and magnetic resonance scans if necessary. Eight tumor types are considered in our study: hepatocellular carcinoma (HCC), intrahepatic cholangiocarcinoma (ICC), metastasis (meta), hepatoblastoma (hepato), hemangioma (heman), focal nodular hyperplasia (FNH), cyst, and others (all other tumor types). If a lesion's type cannot be determined according to image signs~\cite{Marrero2014ACG} and pathology, it will be marked as ``unknown'' and ignored in training and evaluation. In total, 4010 tumor instances are annotated, whose volumes range from 11 to $3.7\times 10^6$ mm$^3$. Detailed statistics and examples of the lesions are shown in the Appendix. We train two separate networks for NC and DCE CTs. In the former setting, both normal and patient data are used and randomly split into 1149 training, 100 validation, and 500 testing. In the latter one, only patient data are used with 641 training, 100 validation, and 200 testing. Another hold-out set of 150 patients and 100 normal CTs are used for reader study to compare our accuracy with two radiologists.

{\bf Implementation Details.} Each CT is resampled to $0.7\times0.7\times5$mm in spacing. We first train an nnU-Net on public datasets to segment liver and surrounding organs (gallbladder, hepatic vein, spleen, stomach, and pancreas), and then crop the liver region to train PLAN. To help PLAN differentiate liver tumors and other organs, we train the network to segment both tumors and organs using the predicted organ labels. PLAN is built on top of the nnU-Net framework~\cite{Isensee2021nnUNet}. Its pixel encoder is a U-Net encoder
, whereas its pixel decoder is a light-weight feature pyramid network~\cite{Cheng2022Mask2Former}
. The lesion branch incorporates three transformer decoder blocks with masked attention~\cite{Cheng2022Mask2Former} which use feature maps of strides 16, 8, 4 from the pixel decoder. The number of random queries is $Q=20$; the embedding dimension is $M=64$; the number of sampled pixels is $K=12544$~\cite{Cheng2022Mask2Former}, foreground pixels $n=3$; the loss weight is 0.1 for the no-object class while 1 for other classes in the lesion branch~\cite{Cheng2022Mask2Former}. 
The weights in Eq.~\ref{eq:loss} are $\lambda_1=\lambda_{2\mathrm{c}}=2, \lambda_{2\mathrm{m}}=5, \lambda_3=1, \lambda_4=0.1$.  We use the RAdam optimizer with an initial learning rate of 0.0001
. Each training batch contains two patches of size $256\times256\times24$. For DCE CT, the three phases form a 3-channel image as the network input. Extensive data augmentation is applied including random cropping, scaling, flipping, elastic deformation, and brightness adjustment~\cite{Isensee2021nnUNet}. During training, we first pretrain the backbone and the pixel branch for 500 epochs, and then train the whole network for another 500 epochs.

\begin{table}[t]
\caption{Patient-level performance on the test set of 500 cases. Spec.~1: specificity on the 202 completely normal cases; Spec.~2: specificity on the 100 hard non-tumor cases
.}
\label{tbl:patient}
\centering
\begin{tabular}{l|p{10mm}<{\centering}p{10mm}<{\centering}p{10mm}<{\centering}|p{15mm}<{\centering}p{10mm}<{\centering}|c}
\hline
            & \multicolumn{3}{c|}{NC tumor screening (\%)}    & \multicolumn{2}{c|}{NC diagnosis AUC} & DCE diagnosis AUC \\
\hline
            & Sens.  & Spec.~1 & Spec.~2 & Malignant        & Benign        & 8-class Average   \\
\hline
nnU-Net~\cite{Isensee2021nnUNet}      & 94.4        & 95.1         & 91.0         & 0.948                 & 0.829              & 0.863          \\
Mask2Former~\cite{Cheng2022Mask2Former} & 93.9        & 97.0         & \bf 94.0         & 0.924                 & 0.828              & 0.873          \\
PLAN (ours)     & \bf 95.0        & \bf 97.5         & \bf 94.0         & \bf 0.961                 & \bf 0.968              & \bf 0.898       \\  
\hline
\end{tabular}
\end{table}

{\bf Patient-level results.} This paper has three major goals: tumor screening in NC CT (classifying a subject as normal or tumor), preliminary diagnosis in NC CT (predicting the existence of malignant and benign tumors), and fine-grained diagnosis in DCE CT (predicting the existence of 8 tumor types). Among the 8 tumor types, HCC, ICC, meta, and hepato are malignant; heman, FNH, and cyst are benign. ``Others'' can be either malignant or benign, thus are excluded in the preliminary diagnosis task. 
The NC test set contains 198 tumor cases, 202 completely normal cases, and 100 ``hard'' non-tumor cases which may have larger image noise, artifact, ascites, diffuse liver diseases such as hepatitis and steatosis. These cases are used to test the robustness of the model in real-world screening scenario with diverse tumor-free images. We compare PLAN with a widely-used strong baseline, nnU-Net~\cite{Isensee2021nnUNet}. The recent mask transformer, Mask2Former~\cite{Cheng2022Mask2Former}, is also adapted to 3D for comparison. For the baselines, patient-level labels are inferred from their predicted masks by counting lesion pixels. As displayed in \Table{patient}, PLAN achieves the best accuracy on all tasks, especially in NC preliminary diagnosis tasks, which demonstrates the effectiveness of its dedicated patient branch that can explicitly aggregate features from the whole image. 

\begin{table}[t]
\caption{Lesion-level performance (precision, recall, recall of lesions with different radius, classification accuracy of 8 tumor types), and pixel-level performance (Dice per case). Precision, recall, and Dice are computed without considering the tumor types.}
\centering
\begin{tabular}{l|l|p{10mm}<{\centering}p{10mm}<{\centering}p{12mm}<{\centering}p{10mm}<{\centering}p{10mm}<{\centering}p{12mm}<{\centering}p{10mm}<{\centering}|p{10mm}<{\centering}}

\hline
&   & Prec. & Recall & R<5mm & 5$\sim$10 & 10$\sim$20 & >20mm & Acc. & Dice \\
\hline
\multirow{3}{*}{NC}  & nnU-Net      & 78.8      & 77.3   & 19.7          & 63.6      & 90.1       & 96.5             & 75.7           & \bf 78.3 \\
                     & Mask2Former & \bf 85.7      & 74.0   & 10.0          & 60.5      & \bf 91.9       & 97.4             & 77.9           & 76.4 \\
                     & PLAN        & 80.1      & \bf 81.9   & \bf 21.9          & \bf 64.6      & 90.1       & \bf 98.3             & \bf 78.5           & 77.2 \\
\hline
\multirow{3}{*}{DCE} & nnU-Net      & 88.1      & 88.3   & 22.5          & \bf 76.4      & 93.7       & \bf 98.3             & 83.1           & \bf 84.2 \\
                     & Mask2Former & 90.3      & 83.5   & 11.7          & 74.4      & \bf 94.6       & 97.4             & 84.8           & 82.9 \\
                     & PLAN        & \bf 92.2      & \bf 89.0   & \bf 25.6          & 74.9      & \bf 94.6       & \bf 98.3             & \bf 85.9           & \bf 84.2 \\
\hline
\end{tabular}
\label{tbl:lesion}
\end{table}

{\bf Lesion and pixel-level results.} In lesion-level evaluation, we treat a prediction as a true positive if its overlap with a ground-truth lesion is >0.2 in Dice. Lesions smaller than 3mm in radius are ignored. As shown in \Table{lesion}, the pixel-level accuracy of nnU-Net and PLAN are comparable, but PLAN's lesion-level accuracy is consistently higher than nnU-Net. In this work, we focus more on patient and lesion-level metrics
. Although NC images have low contrast, they can still be used to segment and classify lesions with $\sim$ 80\% precision, recall, and classification accuracy. It implies the potential of NC CT, which has been understudied in previous works. Mask2Former has higher precision but lower recall in NC CT, especially for small lesions, while PLAN achieves the best recall using the foreground-enhanced sampling loss. Both PLAN and Mask2Former achieve better classification accuracy, which illustrates the mask transformer architecture is good at lesion-level classification.

\makeatletter
\newcommand\figcaption{\def\@captype{figure}\caption}
\newcommand\tabcaption{\def\@captype{table}\caption}
\makeatother

\begin{figure}[t]
\begin{minipage}{0.47\textwidth}
\centering
\begin{tabular}{c|ccp{1cm}<{\centering}|p{1cm}<{\centering}}
\hline
         & \multicolumn{3}{c|}{NC}       & DCE          \\
\hline
         & Sens. & Spec. & 3-class Acc. & 8-class Acc. \\
\hline
Radiologist 1 & 94.1  & \bf 99.0  & 90.8         & \bf 75.6         \\
Radiologist 2 & 85.5  & \bf 99.0  & 72.0         & 40.5         \\
PLAN     & \bf 96.7  & 98.0  & \bf 91.3         & \bf 75.6         \\
\hline
\end{tabular}
\tabcaption{Reader study results on 150 tumor cases and 100 normal cases. 3-class acc.~means classification accuracy of normal vs.~benign vs.~malignant.}
\label{tbl:reader_study}
\end{minipage}
\hspace{2mm}
\begin{minipage}[h]{0.47\linewidth}
\centering
\includegraphics[width=.9\textwidth]{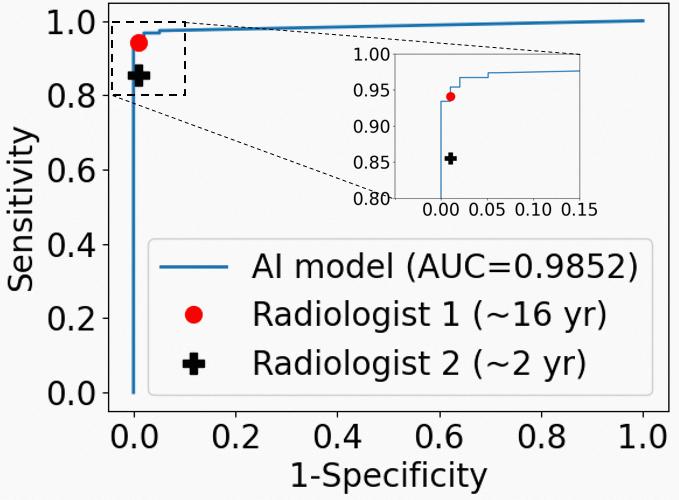}
\figcaption{ROC curve of our method versus 2 radiologists' performance.} 
\label{fig:reader_study}
\end{minipage}%
\end{figure}

{\bf Comparison with radiologists.} In the reader study, we invited a senior radiologist with 16 years of experience in liver imaging, and a junior radiologist with 2 years of experience. They first read the NC CT of all subjects and provided a diagnosis of normal, benign, or malignant. Then, they read the DCE scans and provided a diagnosis of the 8 tumor types. We consider patients with only one tumor type in this study. Their reading process is without time constraint
. In \Table{reader_study} and \Fig{reader_study}, all methods get good specificity probably because the normal subjects are completely healthy. Our model achieves comparable accuracy with the senior radiologist but outperforms the junior one by a large margin in sensitivity and classification accuracy.

\begin{table}[t]
\scriptsize
\centering
\caption{Ablation study on NC data. FES loss: foreground enhanced sampling loss.}
\label{tbl:ablation}
\begin{tabular}{l|p{1.3cm}<{\centering}p{1.cm}<{\centering}|p{1.4cm}<{\centering}p{1cm}<{\centering}|p{1.1cm}<{\centering}p{.9cm}<{\centering}p{.8cm}<{\centering}p{.9cm}<{\centering}}
\hline
            & \multicolumn{2}{c|}{Tumor screening (\%)}    & \multicolumn{2}{c|}{Prelim. diagnosis AUC} & \multicolumn{4}{c}{Lesion and pixel-level (\%)} \\
\hline
                & Sens. & Spec. & Malignant & Benign & Precision & Recall & Acc. & Dice \\
\hline
PLAN (proposed)      & \bf 95.0 & 96.4 & \bf 96.1      & \bf 96.8       & 80.1      & \bf 81.9   & \bf 78.5        & \bf 77.2 \\
w/o anchor queries   & 94.4 & 95.4 & 94.9      & 93.5       & 78.9      & 78.1   & 77.1        & 75.0 \\
w/o FES loss      & 93.4 & 96.0 & 94.0      & 96.4       & \bf 86.6      & 75.1   & 77.7        & \bf 77.2 \\
w/o consistency loss & 93.9     & \bf 96.7     & 95.4          & 96.3           & 79.1      & 80.7   & 78.2        & 76.6   \\
\hline
\end{tabular}
\end{table}

An ablation study for our method is shown in \Table{ablation}. It can be seen that our proposed anchor queries produced by the pixel branch, FES loss, and lesion-patient consistency loss are useful for the final performance. The efficacy of the lesion and patient branches has been analyzed above based on the lesion and patient-level results. We will show the accuracy for each tumor type and more qualitative examples in the Appendix.

{\bf Comparison with literature.} 
In the pixel level, we obtain Dice scores of 77.2\% and 84.2\% using NC and DCE CTs, respectively. The current state of the art (SOTA) of LiTS~\cite{Bilic2019LiTS} achieved 82.2\% in Dice using CTs in venous phase; \cite{Zhang2021modal} achieved 81.3\% in Dice using DCE CT of two phases. In the lesion level, our precision and recall are 80.1\% and 81.9\% for NC CT, 92.2\% and 89.0\% for DCE CT, at 20\% overlap. \cite{Zhou2021Frontiers} achieved 83\% and 93\% for DCE CT. SOTA of LiTS achieved 49.7\% and 46.3\% at 50\% overlap. \cite{Yasaka2018differ} classified lesions into 5 classes, achieving 84\% accuracy for DCE and 49\% for NC CT. We classify lesions into 8 classes with 85.9\% accuracy for DCE and 78.5\% for NC CT. In the patient level, \cite{Cheng2022HCC} achieved AUC=0.75 in NC CT tumor screening, while our AUC is 0.985. In summary, our results are superior or comparable to existing works.

\section{Conclusion}
Three tasks are investigated in this paper: liver tumor screening and preliminary diagnosis in NC CT, and the diagnosis of 8 tumor types in DCE CT. The pixel-lesion-patient network is proposed that can accomplish lesion-level segmentation and classification, and patient-level classification. Comprehensive evaluation on a large-scale dataset confirms the effectiveness and clinical significance of our method.
It can serve as a powerful tool for automated screening and diagnosis of various liver tumors.
Our future work includes further improving the specificity of hard non-tumor cases and sensitivity of small lesions.

\section{Appendix}

\begin{table}[]
	\centering
	\setlength{\tabcolsep}{2pt}
	\caption{Statistics of lesion types and sizes in our dataset of 939 patients and 4010 annotated lesions. The last row shows the number of patients with the tumor type. Note that one patient may have tumors of multiple types.}
	\label{tab:lesion_stats}
	\begin{tabular}{l|cccccccc|c}
		\hline
		& HCC & ICC & Meta & Hepato & Heman & FNH & Cyst & Others & Total \\
		\hline
		R\textless{}5mm    & 19  & 6   & 516  & 3      & 45    & 0   & 1081 & 127    & 1797  \\
		5$\sim$10          & 101 & 4   & 504  & 3      & 118   & 13  & 224  & 45     & 1012  \\
		10$\sim$20         & 186 & 21  & 241  & 1      & 82    & 14  & 43   & 33     & 621   \\
		\textgreater{}20mm & 191 & 59  & 90   & 20     & 138   & 28  & 15   & 39     & 580   \\
		\hline
		Total \# lesions             & 497 & 90  & 1351 & 27     & 383   & 55  & 1363 & 244    & 4010     \\
		\hline
		\# Patients         & 414 & 79  & 218  & 20     & 171   & 43  & 376  & 174    & 939  \\
		\hline
	\end{tabular}
\end{table}

\begin{table}[]
	\centering
	\setlength{\tabcolsep}{0pt}
	\caption{Lesion-level performance (type-agnostic precision, type-agnostic recall, and recall of 8 tumor types). We find recall is negatively correlated with average size of each tumor type. NC: non-contrast; DCE: dynamic contrast-enhanced (NC+arterial+venous).}
	\label{tab:lesion_acc_type}
	\begin{tabular}{l|l|p{10mm}<{\centering}p{10mm}<{\centering}|p{10mm}<{\centering}p{10mm}<{\centering}p{10mm}<{\centering}p{12mm}<{\centering}p{10mm}<{\centering}p{10mm}<{\centering}p{10mm}<{\centering}p{10mm}<{\centering}}
		\hline
		&             & Prec. & Recall & HCC  & ICC   & Meta & Hepato & Heman & FNH   & Cyst & Others \\
		\hline
		\multirow{3}{*}{NC}  & nnUNet      & 78.8      & 77.3   & \bf 85.7 & \bf 100.0 & 60.5 & \bf 66.7   & 84.0  & 80.0  & 76.2 & 65.7   \\
		& Mask2Former & \bf 85.7      & 74.0   & 83.8 & \bf 100.0 & 59.0 & \bf 66.7   & 81.3  & 90.0  & 48.6 & 62.9   \\
		& PLAN        & 80.1      & \bf 81.9   & 82.9 & \bf 100.0 & \bf 63.6 & \bf 66.7   & \bf 85.3  & \bf 100.0 & \bf 77.1 & \bf 80.0   \\
		\hline
		\multirow{3}{*}{DCE} & nnUNet      & 88.1      & 88.3   & \bf 91.4 & \bf 100.0 & 70.8 & \bf 100.0  & \bf 94.7  & \bf 100.0 & 78.0 & \bf 71.4   \\
		& Mask2Former & 90.3      & 83.5   & \bf 91.4 & \bf 100.0 & 65.1 & \bf 100.0  & 86.7  & \bf 100.0 & 56.0 & 68.6   \\
		& PLAN        & \bf 92.2      & \bf 89.0   & 89.5 & \bf 100.0 & \bf 72.8 & \bf 100.0  & 92.0  & \bf 100.0 & \bf 86.2 & \bf 71.4  \\
		\hline
	\end{tabular}
\end{table}


\begin{figure}[]
	\centering
	\includegraphics[width=.6\textwidth,trim=0 0 0 0, clip]{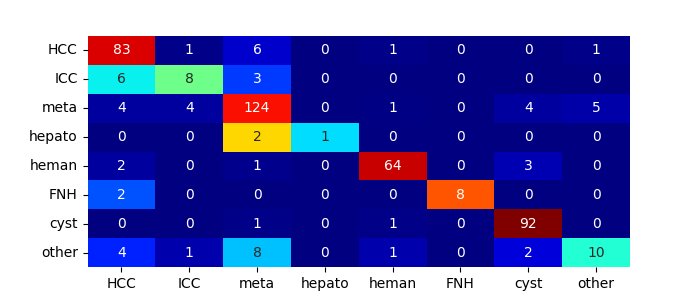}  
	\caption{Confusion matrix of lesion-level tumor classification in DCE CT. Rows and columns indicate true and predicted tumor types, respectively. Only the correctly detected lesions are considered here for evaluation. Cyst, heman, HCC, meta, and FNH can be better classified. 
	ICC and others may be confused with HCC and meta.}
\end{figure}


\begin{figure}[]
\centering
\includegraphics[width=\textwidth,trim=0 50 30 0, clip]{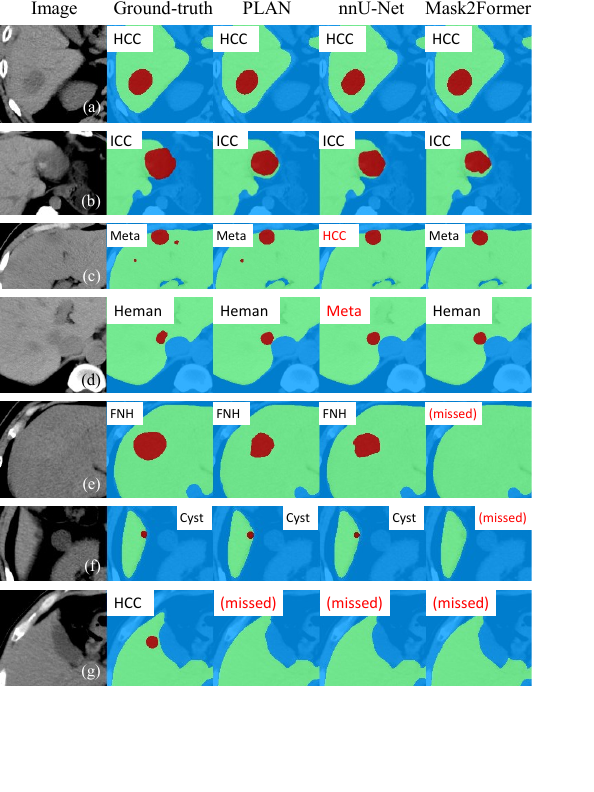}  
\caption{Qualitative examples of lesion segmentation and classification in NC CT using different methods. Ground-truth and predicted tumor types are shown in each result. Ground-truth masks were annotated on registered venous phase. If a tumor is not predicted, it will be marked as ``missed''. Red texts indicate wrong predictions. Some challenging cases can be detected by PLAN, such as the subtle FNH (e), small meta (c) and cyst (f), and the heman which resembles adjacent vessels (d). We also show failure cases in (g) where a very subtle HCC is missed, and in (c) where a tiny meta (among three metas) is missed.}
\end{figure}

%
%
\bibliographystyle{splncs04}
\bibliography{main}

\end{document}